\documentclass{elsart}

\usepackage{psfig}

\usepackage{natbib}

\begin{document}

\runauthor{O'Brien and Bleackley}


\begin{frontmatter}

\title{Coronal Line Emission from NLS1s}

\author[LEI]{P.T. O'Brien, P. Bleackley}
\address[LEI]{Department of Physics \& Astronomy, University of Leicester,
University Road, Leicester, LE1 7RH, U.K.}


\begin{abstract}
We discuss the optical coronal line spectra observed for a sample of 19
Narrow Line Seyfert 1 galaxies. We find no correlation between the
coronal line strength and the soft X-ray power-law index derived from {\it
ROSAT PSPC} data. There is a trend for broader coronal lines to have
larger equivalent widths. In addition, a strong trend is found between
line width and velocity relative to the NLR. This trend is interpreted in
terms of a decelerating outflow, originating close to the nucleus.
\end{abstract}


\end{frontmatter}


\section{Introduction}

As a class Narrow Line Seyfert 1 galaxies (NLS1s) are defined in terms of
their optical emission-line properties \cite{Osterbrock}. In recent
years, however, NLS1s have been studied mainly in relation to their X-ray
emission, which displays some peculiar properties. In particular, the
soft X-ray emission from NLS1s seems to be extreme in its shape and
variability, possibly due to an extreme accretion rate \cite{Pounds}. To
study this emission directly is difficult due to Galactic absorption.
However, an indirect probe is provided by the high-ionization coronal
lines. These forbidden lines arise from species with high ionization
potentials ($>100$ eV), and are thought due to photoionization by the
hard AGN continuum. 

As part of a programme to study their multi-wavelength properties,
we obtained optical spectra of 19 NLS1s with the {\it IDS\/} mounted on
the {\it INT\/} at La Palma. These objects were not chosen to have strong
coronal line emission, but rather to represent the range in observed
X-ray spectral indices as observed using the {\it ROSAT PSPC}. The
optical data were reduced using {\sc STARLINK} software.

\section{X-ray Emission and Coronal Lines}

For each NLS1, we measured the strength and redshift of the strongest 
optical coronal lines, including [Fe VII] $\lambda 6087$, [Fe X] $\lambda
6374$ and [Fe XI] $\lambda 7892$. The Hydrogen and Helium permitted
lines along with several lines emitted from the NLR, including [O III]
$\lambda\lambda 4959, 5007$, were also quantified. 

\begin{figure}[htb]
\centerline{\psfig{figure=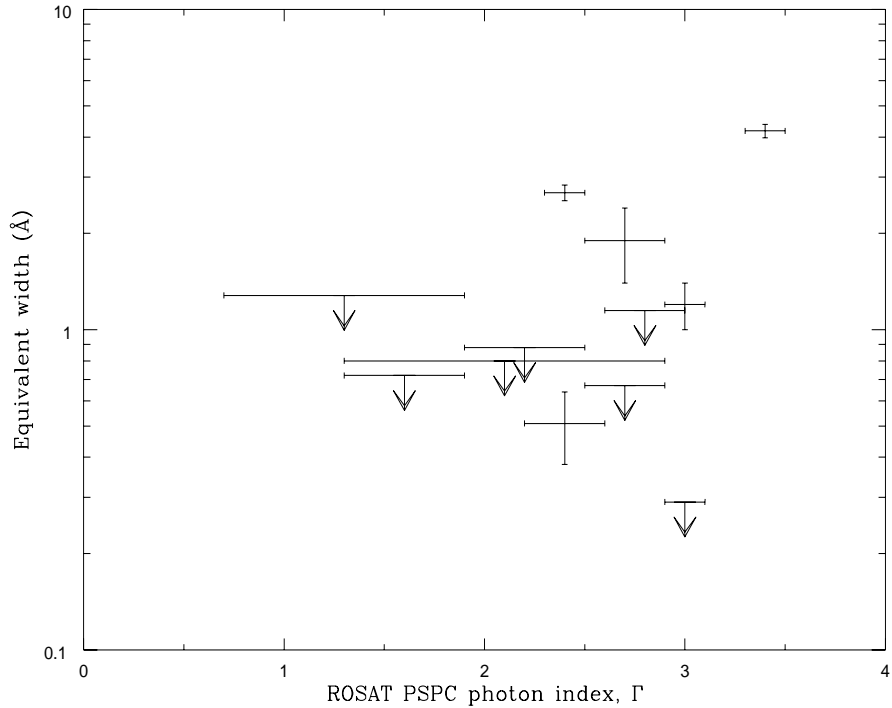,height=4truein,width=5truein,angle=-90}}
\caption{The equivalent width of [Fe X] $\lambda 6374$ versus the
power-law photon  index derived from {\it ROSAT PSPC\/} data.}
\label{fig:feten}
\end{figure}

In a previous study of AGN known to have coronal lines \cite{Erkens}, it
was found that the objects with the largest equivalent widths of
coronal-line emission were those with the steepest soft X-ray spectra,
based on power-law fits to the {\it ROSAT PSPC\/} data. In Figure
\ref{fig:feten} we
show the equivalent width of [Fe X] $\lambda 6374$ versus the photon
index, $\Gamma$ (taken from the literature). No correlation is seen and a
wide range in coronal-line strength is observed for a given power-law
index. 

\section{Kinematics}

\begin{figure}[htb]
\centerline{\psfig{figure=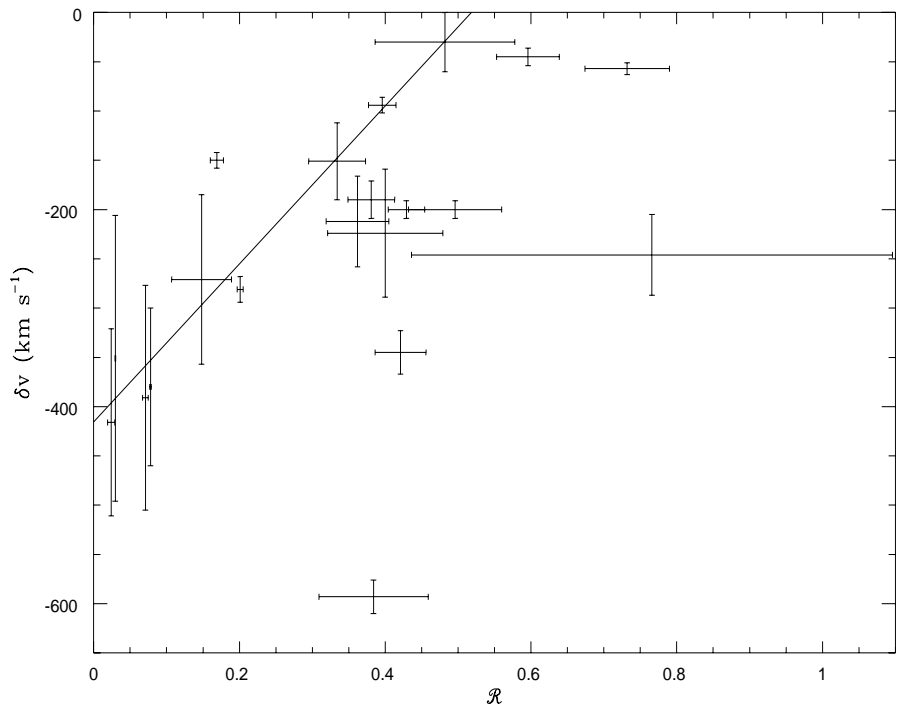,height=4truein,width=5truein,angle=-90}}
\caption{The velocity of the Fe coronal lines relative to [OIII] $\lambda
5007$ versus the radial estimator (defined in the text). A best-fit line
allowing for errors in both variables is also shown.}
\label{fig:outflow}
\end{figure}

The exact location of the coronal line region is unknown, but it is believed
to lie between the BLR and NLR, possibly associated with the dusty torus
invoked in AGN unification models \cite{Nagao}. As resolving the
coronal-line region {\it via} direct imaging is difficult, to place some
constraints we measured the FWHM and redshift of the coronal lines
relative to the strong NLR line, [O III] $\lambda 5007$. Assuming the
line widths are governed primarily by orbital motion, we define a `radial
estimator', $R$, such that
\begin{displaymath}
R = \frac{r_{[\mathrm{Fe}]}}{r_{[\mathrm{OIII}]}} =
\left(\frac{\mathrm{FWHM}_{[\mathrm{OIII}]}}{\mathrm{FWHM}_{\mathrm{[Fe}]}}\right)^2 ,
\end{displaymath}
where $r_{[\mathrm{Fe}]}$ and $r_{[\mathrm{OIII}]}$ are the radial
distances from the centre. In Figure \ref{fig:outflow} we plot the
velocity of the coronal line (relative to [O III] $\lambda 5007$) versus
the radial estimator. A clear correlation is present such that broader
coronal lines have larger blueshifts. The simplest interpretation is that
the coronal-line gas is part of a decelerating outflow. The broader
coronal lines also tend to have larger equivalent widths.

\section{Conclusions}

The lack of correlation between the coronal-line strengths and the {\it
ROSAT} power-law indices is somewhat surprising given the predictions of
photoionization models. It may be that the range in gas conditions or
covering factors for the coronal-line regions between different NLS1s is
quite large. The continuum shape may also be aspect-dependent, such that
the steep soft X-ray continuum is not always seen by the coronal-line
emitting gas. Finally, a single power law may provide a poor
parameterization of the soft X-ray spectral shape. {\it ASCA} spectra of
NLS1s do suggest a complicated spectral shape in some objects
\cite{Vaughan}. The shape of the soft X-ray continuum in NLS1s will be
accurately determined by forthcoming {\it Chandra\/} and {\it
XMM-Newton\/} observations.

The kinematical results suggest a connection between the coronal-line
emitting gas and the central region of NLS1s. In the context of a
decelerating outflow model, the gas velocity is $\approx 500$ km s$^{-1}$ at
a distance from the centre $\approx 1/40$ that of the NLR. Given an NLR size
$\approx 100$~pc, this implies that the coronal-line gas originates at
$\approx 2$~pc from the centre. Such a small size could be associated
with the outer region of the BLR and/or the inner edge of the proposed
dusty torus.



\end{document}